\documentclass[11pt,a4paper]{article}
\usepackage[T1]{fontenc}

\usepackage{graphicx}
\usepackage{bm}
\usepackage{hyperref}

\usepackage{amsmath,amsfonts,amssymb}

\begin{document}
\begin{titlepage}
\begin{flushright}
    { IPMU18-0020}
\end{flushright}
\vspace{1cm}

\begin{flushleft}
    \textbf{\huge Cosmologically allowed regions for\\[0.6em]
the axion decay constant $F_a$}
\end{flushleft}

\vspace{2cm}

\noindent
\textbf{\Large Masahiro Kawasaki}$^{a,b}$
\textbf{\Large Eisuke Sonomoto}$^{a,b}$
\textbf{\Large Tsutomu T. Yanagida}$^{b,c}$

\vspace{0.5cm}

\noindent
$^{a}$ICRR, The University of Tokyo, Kashiwa, Chiba 277-8582, Japan\\[0.3em]
$^{b}$Kavli IPMU (WPI), UTIAS, The University of Tokyo, Kashiwa, Chiba 277-8583, Japan\\[0.3em]
$^{c}$Hamamatsu Professor

\vspace{0.3cm}
\noindent
{\small 
Email: kawasaki@icrr.u-tokyo.ac.jp,
sonomoto@icrr.u-tokyo.ac.jp,
tsutomu@ipmu.u-tokyo.ac.jp
}

\vspace{3cm}

\abstract{
If the Peccei-Quinn symmetry is already broken during inflation, the decay constant $F_a$ of the axion can be in a wide region from $10^{11}$~GeV to $10^{18}$~GeV for the axion being the dominant dark matter. In this case, however, the axion causes the serious cosmological problem, isocurvature perturbation problem, which severely constrains the Hubble parameter during inflation. To avoid this problem, Peccei-Quinn scalar field must take a large value $\sim M_{p}$ (Planck scale) during inflation.  
In this letter, we point out that the allowed region of the decay constant $F_a$ is reduced to a rather narrow region for a given tensor-to-scalar ratio $r$ when Peccei-Quinn scalar field takes  $\sim M_{p}$ during inflation.
For example, if the ratio $r$ is determined as $r \gtrsim 10^{-3}$ in future measurements, we can predict $F_a \simeq (0.1-1.4)\times 10^{12}$~GeV for domain wall number $N_\text{DW}=6$.
}

\end{titlepage}


\section{Introduction}

The QCD axion~\cite{Weinberg:1977ma,Wilczek:1977pj,Kim:1979if,Shifman:1979if,Dine:1981rt} predicted in the Peccei-Quinn(PQ) mechanism~\cite{Peccei:1977hh,Peccei:1977ur} is very interesting, since it not only solves the strong CP problem in QCD, but also provides the dark matter density in the present universe~\cite{Preskill:1982cy,Abbott:1982af,Dine:1982ah}.
Thus, the experimental detection of the dark matter axion is very important.
The detection rate, however, depends on the axion decay constant $F_a$.
Therefore, the theoretical determination of the decay constant is very crucial to test the dark matter axion hypothesis.

If the PQ symmetry is broken down after inflation, it produces too many domain walls~\cite{Sikivie:1982qv} and most of the axion models are excluded except for the model with $N_\text{DW} =1$ ($N_\text{DW}$: domain wall number or color anomaly factor).
For the case of $N_\text{DW} =1$, the produced domain walls are not stable and their collapsion produces many axions.
The detailed lattice simulations~\cite{Hiramatsu:2012gg,Kawasaki:2017kkr} have shown that the resultant axions can provide us with the observed DM density if $F_a\simeq 5\times 10^{10}$~GeV.

On the other hand, if the PQ symmetry is already broken during inflation, it is known that $F_a$ can take a large region from $10^{11}$~GeV to $10^{18}$~GeV.
In this case, however, the axion field acquires quantum fluctuations with amplitude $\simeq H_\text{inf}/2\pi$ ($H_{\rm inf}$: the Hubble parameter during inflation), which result in isocurvature perturbations of the cold dark matter (CDM)~\cite{Axenides:1983hj,Seckel:1985tj,Linde:1985yf,Linde:1990yj,Turner:1990uz}.
Since the isocurvature perturbations are stringently constrained by observations of the cosmic microwave background (CMB)~\cite{Ade:2015lrj}, $H_{\rm inf}$ should be very small, e.g. $H_{\rm inf} \lesssim 10^8$ GeV. However, this problem is relaxed when the PQ field takes a large value like Planck scale ($M_p$) during inflation as first Linde pointed out~\cite{Linde:1991km}. Because the tensor mode (gravitational waves) produced by inflation is directly related to $H_\text{inf}$, we obtain the constraint of $F_a$ and the tensor-to-scalar ratio $r$ from the isocurvature perturbations.
In this letter, we focus on the case of the PQ symmetry broken during inflation and the PQ field taking $M_p$ during inflation, and consider the isocurvature perturbation constraint.
Taking into account the dynamics of the PQ scalar, we show that for a given $r$ the axion can be the dominant dark matter for a rather narrow range of $F_a$.
In the near future the tensor mode with $r\gtrsim 10^{-3}$ can be detected~\cite{Matsumura:2013aja}, so if $r$ is determined we can predict the axion decay constant $F_a$.
For example, $F_a$ should be $(0.1-1.4)\times 10^{12}$~GeV for $r=0.001$ and $N_\text{DW}=6$.

In the next section we study the isocurvature perturbations in axion models.
Concrete axion models which satisfy the isocurvature perturbation constraint and account for the dark matter are described in Sec.~\ref{sec:concrete_models}. Sec.\ref{sec:conclusion} is devoted to conclusion.

\section{Isocuravture perturbations}

During inflation the axion field acquires fluctuations whose amplitude is given by
\begin{equation}
   \delta a \simeq \frac{H_\text{inf}}{2\pi} ,
\end{equation}
where $H_\text{inf}$ is the Hubble parameter during inflation.
This leads to the fluctuation of the misalignment angle,
\begin{equation}
   \delta \theta_a = \frac{\delta a}{F_a}
   = \frac{H_\text{inf}}{2\pi F_a}.
   \label{eq:delta_theta}
\end{equation}
Here $F_a$ is related to the present vacuum expectation value $v$ of the PQ scalar $\Phi$ as
\begin{equation}
   F_a = \frac{v}{N_\text{DW}}.
\end{equation}
Also notice that the phase of the PQ field $\theta_\text{PQ}$ is written as $\theta_\text{PQ}= \theta_a/N_\text{DW}$.

After the axion acquires a mass at the QCD scale, the fluctuations of the axion field induce the axion density perturbations $\delta\rho_a$ which contribute to CDM isocurvature perturbations $S_\text{CDM}$ as
\begin{equation}
   S_\text{CDM}=\frac{\delta\rho_\text{CDM}}{\rho_\text{CDM}}
   =\frac{\Omega_a}{\Omega_\text{CDM}}\frac{\delta\rho_a}{\rho_a},
  \label{eq:cdm_iso}
\end{equation}
where $\Omega_{a\, \text{(CDM)}}$ is the axion (CDM) density parameter and $\rho_{a\,\text{(CDM)}}$ is the axion (CDM) density.
The CDM isocuravture perturbations are stringently constrained by the CMB observations as~\cite{Ade:2015lrj}
\begin{equation}
   \beta_\text{iso} \equiv
   \frac{\mathcal{P}_\text{iso}(k_*)}{
   \mathcal{P}_\text{iso}(k_*)+\mathcal{P}_\text{adi}(k_*)}
   < 0.038 ,
   \label{eq:iso_obs_const}
\end{equation}
where $\mathcal{P}_\text{iso}(k_*)$ and $\mathcal{P}_\text{adi}(k_*)~(\simeq 2.2\times 10^{-9})$ are power spectra of isocuravture and adiabatic perturbations at the pivot scale $k_* =0.05$~Mpc$^{-1}$.

Hereafter, we assume that the axion is the dark matter of the universe, i.e., $\Omega_a = \Omega_\text{CDM}$.
The cosmic axion density is given by
\begin{equation}
   \Omega_a h^2 \simeq 0.18 \, \theta_a^2
   \left(\frac{F_a}{10^{12}\text{GeV}}\right)^{1.19},
\end{equation}
where $h$ is the present Hubble parameter in units of $100$km/sec/Mpc.
Requiring $\Omega _ah^2\simeq \Omega_\text{CDM}h^2\simeq 0.12$~\cite{Ade:2015lrj}, we obtain
\begin{equation}
   \theta_a \simeq 0.82 \,
   \left(\frac{F_a}{10^{12}\text{GeV}}\right)^{-0.595} .
   \label{eq:theta_DM_axion}
\end{equation}
In this case, the fluctuation of the misalignment angle should be less than $\theta_a$, otherwise the axion isocurvature perturbations are too large.
Then, the axion density perturbations are $\delta\rho_a/\rho_a\simeq 2\delta\theta_a/\theta_a$ and Eq.~(\ref{eq:cdm_iso}) is written as
\begin{equation}
   S_\text{CDM}=\frac{2\delta\theta_a}{ \theta_a}.
   \label{eq:cdm_iso_axion_DM}
\end{equation}
With use of Eqs.~(\ref{eq:delta_theta}) and (\ref{eq:theta_DM_axion}) the power spectrum of the CDM isocurvature perturbations is given by
\begin{equation}
   \mathcal{P}_\text{iso} =\langle |S_\text{CDM}|^2 \rangle
    = \left(\frac{H_\text{inf}}{\pi F_a\theta_a}\right)^2.
   \label{eq:iso_power_spec}
\end{equation}
From the constraint on isocurvature perturbations Eq.~(\ref{eq:iso_obs_const})  we obtain
\begin{equation}
   H_\text{inf} \lesssim 2.4\times 10^7~\text{GeV}\,
   \left(\frac{F_a}{10^{12}\, \text{GeV}}\right)^{0.405}.
   \label{eq:hubble_const}
\end{equation}
The tensor perturbation (gravitational waves) produced during inflation is directly related to the Hubble parameter as
\begin{equation}
   r = 1.6\times 10^{-5}\,\left(\frac{H_\text{inf}}{10^{12}\, \text{GeV}}\right)^2 .
\end{equation}
Thus, the dark matter axion seems to predict too small $r$ to be detected in the near future.

\begin{figure}[t]
  \centering \includegraphics[width=7.6cm]{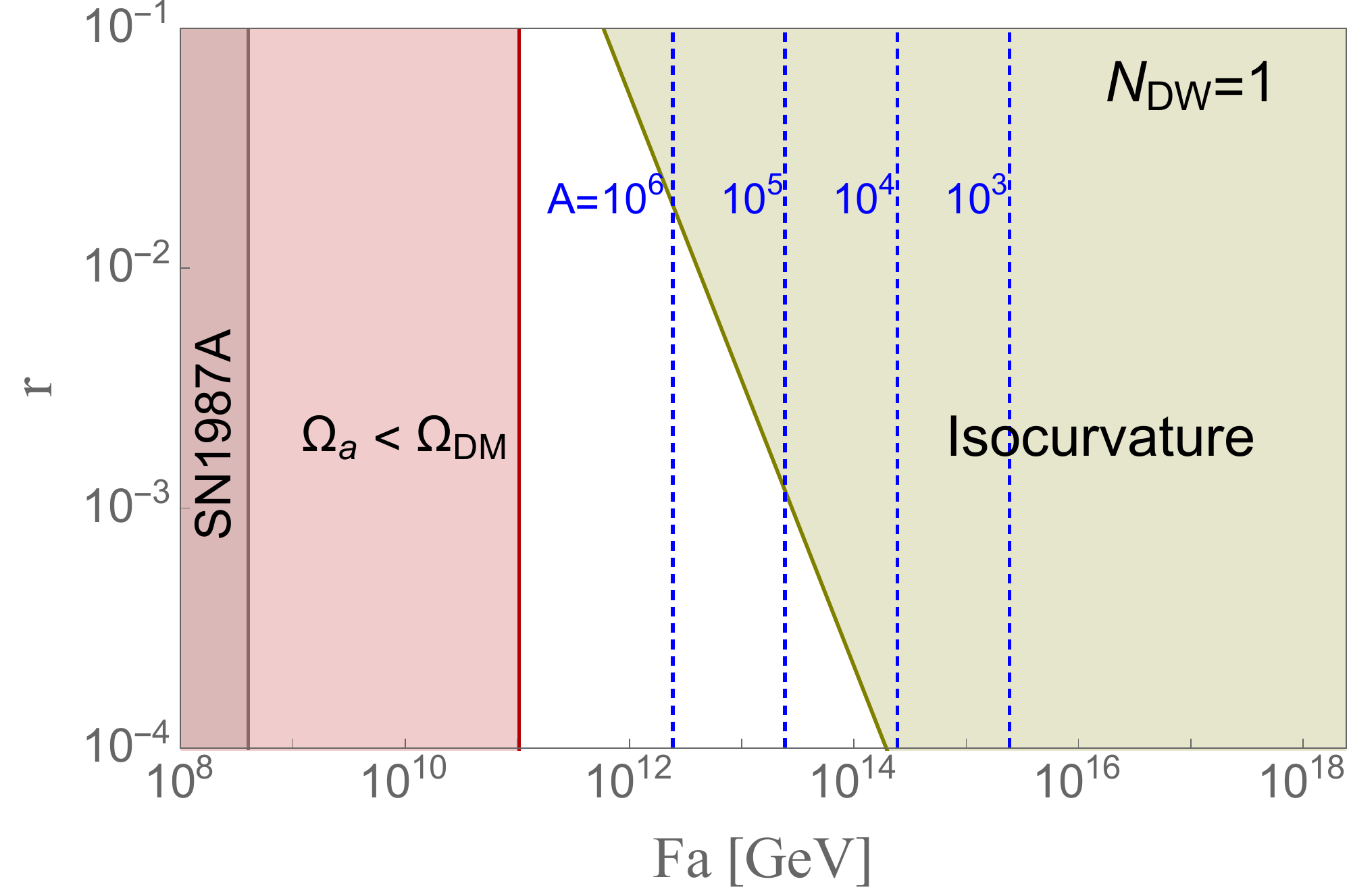}
  \includegraphics[width=7.6cm]{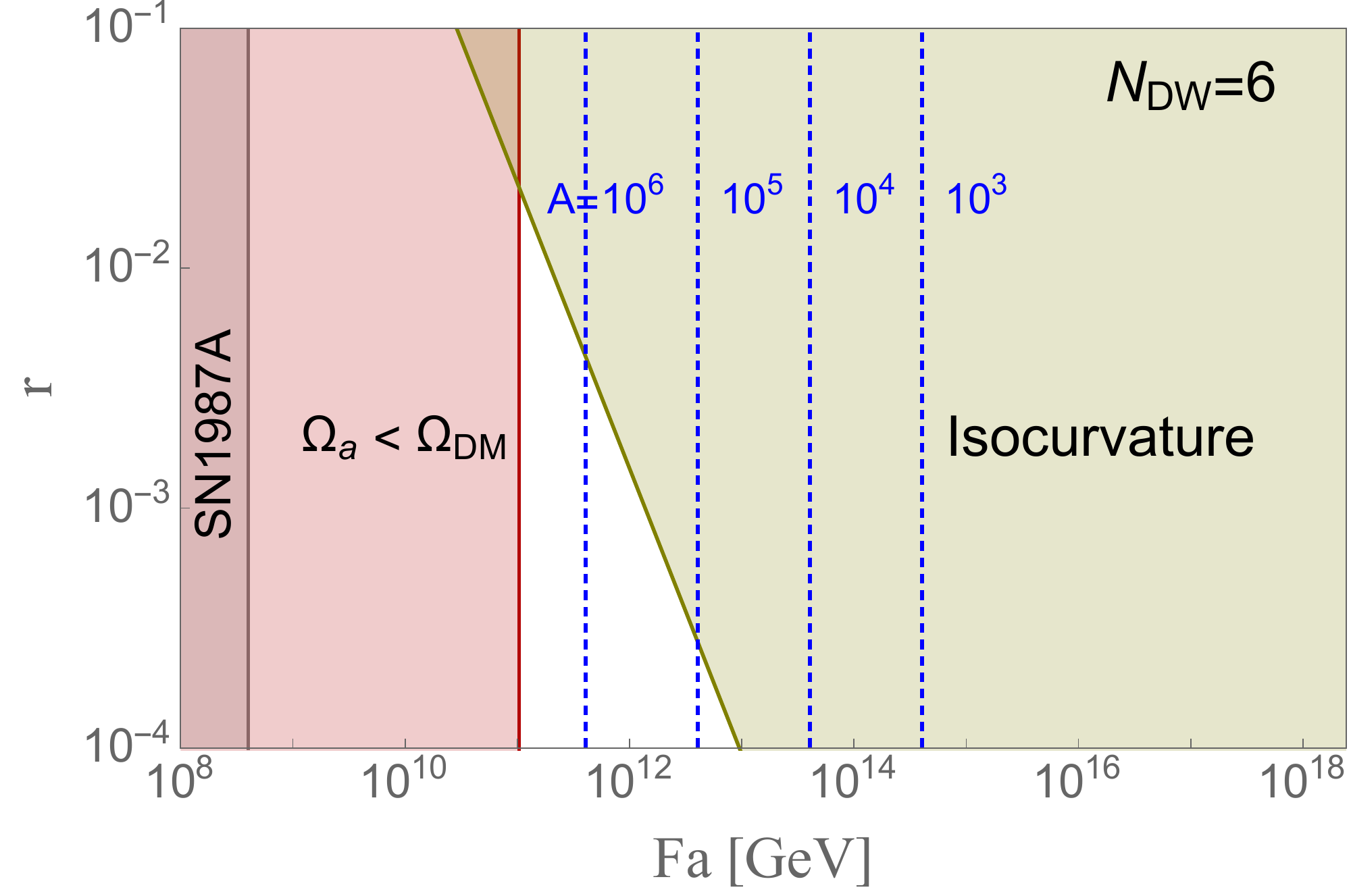}
  \caption{
  Constraints on the tensor-to-scalar ratio $r$ and the axion decay constant $F_a$ for $N_\text{DW}=1$(left) and $N_\text{DW}=6$(right).
  We assume that the expectation value of the PQ field during inflation is $M_p$.
  The dark yellow regions are excluded by the observation of isocurvature perturbations and the red regions are excluded because the axion cannot account for all dark matter. $F_a < 4\times 10^8$~GeV is excluded by SN1987A~\cite{Raffelt:2006cw}.
  We also plot the contours for $A = M_p/v$.}
  \label{fig:constraint_on_r}
\end{figure}

However, the above constraints only apply when the PQ field $\Phi$ has the vacuum expectation value $v (= F_a N_\text{DW})$ during inflation.
As first pointed out by Linde~\cite{Linde:1991km,Kasuya:1996ns,Kawasaki:2013iha,Chun:2014xva,Choi:2014uaa}, when the PQ field has a larger expectation value, the isocurvature perturbations are suppressed.
The constraints are the most relaxed when the PQ field takes the Planck scale $M_p (\simeq 2.4\times 10^{18}~\text{GeV})$ during inflation.
In this case the fluctuation of the misalignment angle is obtained from Eq.~(\ref{eq:delta_theta}) with $F_a$ replaced with $M_p/N_\text{DW}$ as
\begin{equation}
   \delta \theta_a = \frac{N_\text{DW} H_\text{inf}}{2\pi M_p}
   \label{eq:delta_theta_Mp}
\end{equation}
This gives the constraint on the Hubble parameter during inflation,
\begin{equation}
   H_\text{inf} \lesssim 5.7\times 10^{13}\,\text{GeV}
   \frac{1}{N_\text{DW}}
   \left(\frac{F_a}{10^{12}\, \text{GeV}}\right)^{-0.595} ,
\end{equation}
and the tensor-to-scalar ratio,
\begin{equation}
   r \lesssim 0.051\, \frac{1}{N_\text{DW}^2}
   \left(\frac{F_a}{10^{12}\, \text{GeV}}\right)^{-1.19}.
\end{equation}
The constraint from the isocurvature perturbations is shown by dark yellow regions in Fig.~\ref{fig:constraint_on_r} for $N_\text{DW}=1$ and $N_\text{DW}=6$.
Since we assume that the axion accounts for all dark matter, Eq.~(\ref{eq:theta_DM_axion}) should be satisfied for $\theta_a < \pi$, which leads to the constraint,
\begin{equation}
   F_a \gtrsim 1.1\times 10^{11}~\text{GeV}.
\end{equation}
This constraint is shown in red shaded regions in Fig.~\ref{fig:constraint_on_r}.
From the figure it is seen that dark matter axion is consistent with high scale inflation models with the tensor-to-scalar ratio $r$ as large as $0.1$ (0.02) for $N_\text{DW}=1 (6)$. (Notice that the present observational constraint on the tensor mode is $r\lesssim 0.1$~\cite{Ade:2015lrj}.)
Such large $r$ can be probed in future CMB experiments~\cite{Wu:2016hul,Suzuki:2015zzg,Matsumura:2013aja}.
Furthermore, for a given $r$ we obtain a constraint on $F_a$ as
\begin{align}
   1.1\times 10^{11}\text{GeV} ~\lesssim F_a  ~\lesssim
   4.0\times 10^{12}\text{GeV}~N_\text{DW}^{-1.68}~\left(\frac{r}{0.01}\right)^{-0.84},
\end{align}
which provides useful information to the axion dark matter search.

\section{Concrete models}\label{sec:concrete_models}

In this section we show some examples of axion models in which the PQ field has a large expectation value during inflation and hence isocurvature perturbations are suppressed.

\subsection{Quartic potential}

The first model is the standard potential with a quartic term,
\begin{equation}
   V(\Phi) = -m^2 |\Phi|^2 +\frac{1}{2}\lambda |\Phi |^4
   + c_H H^2 |\Phi |^2,
   \label{eq:PQ_potential}
\end{equation}
where $-m^2$ is the mass squared at the origin and $\lambda$ is the coupling constant.
Here we have added the so-called Hubble induced mass term which comes from the interaction with the inflaton like $\mathcal{L}_\text{int} = - (c_H/3)V_\text{inf} |\Phi|^2$ ($V_\text{inf}$: inflaton potential).
The coefficient $c_H$ is assumed to be negative.
Without the Hubble induced mass term the potential has a minimum at $|\Phi | = v=m/\sqrt{\lambda}$.
During inflation the PQ field takes $|\Phi | = \sqrt{-c_H/\lambda}H_\text{inf}$ which can be as large as $M_p$ if $\lambda$ is sufficiently small ($\sim (H_\text{inf}/M_p)^2)$.\footnote{
Even in the absence of  the Hubble induced mass term, the PQ field can have a large expectation value $\sim M_p$ during inflation if the potential is flat enough ($\lambda < (H_\text{inf}/M_p)^2$).}

However, this model has a problem.
After inflation, the Hubble parameter changes rather quickly.
Then, the quartic term causes oscillation of the PQ field and fluctuations of the PQ field increase through parametric resonance, from which the PQ field potential Eq.(\ref{eq:PQ_potential}) obtains a correction due to the fluctuations given by $\Delta V \simeq \lambda \langle |\delta\Phi |^2\rangle |\Phi |^2$.
Therefore if the fluctuation grows as large as $\langle |\delta\Phi |^2\rangle \sim v^2$,  $U(1)_\text{PQ}$ symmetry is restored non-thermally~\cite{Kofman:1995fi}\footnote{
Even if $U(1)_\text{PQ}$ symmetry is not restored, domain walls may be formed because the PQ field settles down to the two different minima, e.g. $v$ and $-v$, due to its initial fluctuations made by inflation. However, this constraint is much weaker than that of parametric resonance~\cite{Kasuya:1997td}.
}.
Once $U(1)_\text{PQ}$ is restored, strings and domain walls are formed and we confront the serious domain wall problem.
In order to avoid the problem, the PQ field should settle down to the minimum before the fluctuations fully develop.
This leads to a constraint on the ratio $A$ of the initial field value after inflation $\Phi_i$ to the breaking scale $v$ as~\cite{Kawasaki:2013iha,Kasuya:1998td,Kasuya:1999hy}
\begin{equation}
   A \equiv \frac{|\Phi_i|}{v} = \frac{|\Phi_i|}{N_\text{DW} F_a}\lesssim 10^4.
   \label{eq:no_DW_condition}
\end{equation}
For $|\Phi_i|\simeq M_p$, as seen in Fig.~\ref{fig:constraint_on_r}, this constraint leads to $r < 10^{-4}$, which is too small to be detected.

\subsection{Sextet potential}

The above problem is solved if the oscillation is controlled by higher order terms like $\Phi^6, \Phi^8, \ldots$.
Let us consider the following potential~\cite{Harigaya:2015hha}:
 \begin{equation}
    V(\Phi) = -m^2 |\Phi|^2 +\frac{1}{2}\lambda |\Phi |^4
    + \frac{1}{3}\eta \frac{|\Phi|^6}{M_p^2} + c_H H^2 |\Phi |^2,
 \end{equation}
 where $\eta$ is the coupling constant.
Similarly to the previous model if $\eta$ is sufficiently small ($\eta \sim (H_\text{inf}/M_p)^2$), the PQ field takes $|\Phi|\sim M_p$ during inflation.
After inflation the PQ field starts to oscillate under $\Phi^6$ potential until the amplitude becomes smaller than $\phi_4$ given by
\begin{equation}
   \phi_4 \simeq \sqrt{\frac{\lambda}{\eta}}M_p.
\end{equation}
During this period the parametric resonance is ineffective because the oscillation due to the sextet term is slow enough.
Afterwards, the oscillation is caused by the quartic term and the parametric resonance takes place.
However, from Eq.(\ref{eq:no_DW_condition}), if $\phi_4 / v \lesssim 10^4$, the domain walls are not formed, which gives a constraint
\begin{equation}
   v = N_\text{DW}F_a > 10^{-4} \sqrt{\frac{\lambda}{\eta}}M_p.
\end{equation}
This is satisfied e.g. for $\eta \sim 10^{-10}$, $\lambda \sim 10^{-16}$, $H_\text{inf} \sim 10^{13}$~GeV, $m\sim 10$~TeV and $v\sim 10^{12}$~GeV.
This model can be realized also in the framework of supersymmetry(SUSY)~\cite{Harigaya:2015hha}.

\subsection{SUSY axion model}

Another interesting model is the SUSY axion model which is described by the following superpotential:
\begin{equation}
   W = h \Phi_0 (\Phi_{+}\Phi_{-} - v^2),
   \label{eq:susy_axion_superpot}
\end{equation}
where superfields $\Phi_0$, $\Phi_{+}$ and $\Phi_{-}$ have PQ charges (R charge) $0 (+2)$, $+1 (0)$ and $-1 (0)$, respectively.
From Eq.~(\ref{eq:susy_axion_superpot}), the scalar potential is
\begin{equation}
   V_\text{SUSY} = h^2|\Phi_{+} \Phi_{-} -v^2|^2
   + |\Phi_0|^2 (|\Phi_{+}|^2+|\Phi_{-}|^2),
\end{equation}
which has a flat direction,
\begin{equation}
   \Phi_{+}\Phi_{-} = v^2,~~~~~~\Phi_0=0.
\end{equation}
If we include soft SUSY breaking mass terms, the scalar potential is written as
\begin{equation}
   V = V_\text{SUSY} + m_{+}^2|\Phi_{+}|^2 + m_{-}^2|\Phi_{-}|^2
   + m_{0}^2|\Phi_{0}|^2,
\end{equation}
where $m_{\pm, 0}$ are soft masses $\sim \mathcal{O}(1)$~TeV.
This potential has the minimum at
\begin{align}
   |\Phi_{+}| & \equiv \phi_{+} =\sqrt{\frac{m_{-}}{m_{+}}} v \\
   |\Phi_{-}| & \equiv \phi_{-} = \sqrt{\frac{m_{+}}{m_{-}}} v \\
   |\Phi_0| & = 0 .
\end{align}
where $\phi_\pm$ are radial components of $\Phi_\pm$.
The axion is identified as
\begin{equation}
   a = \frac{1}{\sqrt{\phi_{+}^2 + \phi_{-}^2}}
   \left(\phi_{+}^2 \theta_{+} - \phi_{-}^2 \theta_{-}\right) ,
\end{equation}
where $\theta_{\pm}$ are phases of $\Phi_{\pm}$.

In general the Hubble induced mass terms appear in SUSY models,
so during inflation the scalar potential is written as
\begin{equation}
   V =  V_\text{SUSY} + c_{+}H^2|\Phi_{+}|^2 ++ c_{-}H^2|\Phi_{-}|^2
   + c_{0}H^2|\Phi_{0}|^2,
\end{equation}
where $c_{\pm, 0}$  are coefficients of $\mathcal{O}(1)$.
When $c_{+} < 0$, $c_{-} >0$ and $c_0 >0$ during inflation, for example, the scalar fields settle down to the flat direction satisfying $|\Phi_{+}| \simeq M_p \gg |\Phi_{-}|\simeq v^2/M_p$.
In this case the axion field is almost identical to the phase $\theta_{+}$ of $\Phi_{+}$ (more precisely $a\simeq \phi_{+}\theta_{+}$).\footnote{
When $c_{-}<0$ and $c_{+}>0$ the argument applies with interchanging subscripts $+$ and $-$.
Furthermore, if $c_{+} <0$ and $c_{-}<0$, $|\Phi_{+}| \gg |\Phi_{-}|$ or $|\Phi_{+}|\ll |\Phi_{-}|$ is realized depending on their initial values.}
Since the phase perpendicular to the axion is roughly proportional to $\theta_{+}+\theta_{-}$ and has a huge mass $\sim \phi_{+}$, the phase fluctuations satisfy $\delta\theta_{+}\simeq -\delta\theta_{-}$.
Taking into account that $\delta a \simeq \phi_{+}\delta\theta_{+} \simeq H_\text{inf}/2\pi$, we obtain the fluctuation of the misalignment angle as
\begin{equation}
   \delta\theta_a = \frac{N_\text{DW}}{\phi_{+}^2 + \phi_{-}^2}
   (\phi_{+}^2\delta\theta_{+} - \phi_{-}^2\delta\theta_{-})
   \simeq \frac{N_\text{DW} H_\text{inf}}{2\pi \phi_{+}}
   \simeq \frac{N_\text{DW} H_\text{inf}}{2\pi M_p}.
\end{equation}
This is the same as Eq.~(\ref{eq:delta_theta_Mp}) and the isocurvature perturbations are suppressed.

Although $\Phi_{+}$ and $\Phi_{-}$ start to oscillate and their fluctuations are produced through parametric resonance~\cite{Kasuya:1996ns}, non-thermal restoration of $U(1)_\text{PQ}$ does not take place because the flat direction is preserved.
However, domain walls could be formed if the axion fluctuations were so large that the distribution of the misalignment angle became completely random.\footnote{
In this case, since $U(1)_\text{PQ}$ is not restored, produced domain walls have no boundaries.}
The recent lattice simulation~\cite{Kawasaki:2017kkr} shows that the misalignment angle distribution has a peak at the initial value and domain walls are not formed at least for $A < 10^3$.
The result of the simulation and the fact that $U(1)_\text{PQ}$ is not restored suggest that the domain wall formation is unlikely in this model even for $A > 10^{3}$.

\section{Conclusions}
\label{sec:conclusion}

We have studied the axion dark matter hypothesis and derived constraints on the axion decay constant $F_a$ and the tensor-to-scalar ratio $r$ from consideration of the axion isocurvature perturbations.
In deriving the constraints we have assumed that the PQ scalar has a large expectation value $\sim M_p$ during inflation, otherwise we have severer constraints and almost all region of $F_a$ is excluded for $r \gtrsim 10^{-3}$ [see Eq.~(\ref{eq:hubble_const})].
The tensor mode can be detected by observing the CMB B-mode polarization and future experiments on the earth (satellite experiments) can probe the $r \gtrsim 0.01$ ($r \gtrsim 0.001$).
We have shown that if $r$ is detected the axion can account for all dark matter in a rather narrow range of $F_a$.
For example, if a measurement determines $r=0.01$ we can predict  $F_a \simeq (1.1-2.0)\times 10^{11}\,\text{GeV}$ for $N_\text{DW}=6$.
For the smallest $r (\simeq 0.001)$ that can be detected in the near future, $F_a$ should be $\simeq (0.1-1.4)\times 10^{12}$~GeV for $N_\text{DW}=6$.
Therefore, measurements of the tensor mode produced during inflation provide important information for the dark matter axion search experiments.
Finally this result assumes the existence of the axion models where the PQ field has a large expectation value $\sim M_p$ during inflation.
We have shown concrete examples of such axion models.

\section*{Acknowledgements}

This work is supported by Grant-in-Aid for Scientific Research from the Ministry of Education, Science, Sports, and Culture (MEXT), Japan, No. 15H05889 (M.K.), No. 17H01131 (M.K.), 17K05434 (M.K.), No. 26104009 (T.T.Y.), No. 26287039 (T.T.Y.) and No. 16H02176 (T.T.Y.), and
World Premier International Research Center Initiative (WPI Initiative), MEXT, Japan.


\bibliographystyle{JHEP}
\bibliography{axion_iso}

\end{document}